\documentclass[11pt]{article}

\usepackage{palatino}

\usepackage{latexsym}
\usepackage{amssymb}
\usepackage{amsmath, graphicx} 

\usepackage{float}
\restylefloat{figure}

\newtheorem{theorem}{Theorem}		
\newtheorem{corollary}{Corollary}

\newtheorem{proposition}[theorem]{Proposition}   

\newtheorem{definition}[theorem]{Definition}    
\newtheorem{remark}[theorem]{Remark}		

\numberwithin{equation}{section}
\numberwithin{theorem}{section}

\newcommand{\backto}{\,\vrule height6pt width.4pt depth0pt
        \vrule width6pt height.1pt depth0pt\,}    

\font\caps=cmcsc10

\def\End{{\rm End\, }}
\def\Tr{{\rm Tr\, }}
\def\Ker{{\rm Ker\, }}

\def\ad{{\rm ad\, }}

\begin{document}

\title{\Large \bf Traces, symmetric functions, and a raising operator:  \\
      \large  a generating function for Cauchy's enumeration formula}

\author{Jerzy Kocik \\
 \small  Department of Mathematics, Southern Illinois University, Carbondale, IL 62901, \\
  \small \it jkocik@siu.edu}

\date{}
\maketitle

\begin{abstract} 
The polynomial relationship between elementary symmetric functions
(Cauchy enumeration formula) is formulated via a ``raising operator"
and Fock space construction.  
A simple graphical proof of this relation is proposed.
The new operator extends the Heisenberg algebra so that 
the number operator becomes a Lie product.
This study is motivated by natural appearance of these polynomials in the
theory of invariants for Lax equations and in classical and topological
field theories.  

\smallskip
\noindent

\end{abstract}


%
%



\section{Motivations}  \label{s:1}

An endomorphism of a linear space $L$ may be characterized by its 
invariants with respect to some symmetry group, most generally
$GL(n)$,  $n=\dim L$.  The basic invariants are the determinant 
and the trace, which are special cases of the coefficients of 
the secular equations (here called prodeterminants), but also include 
power traces.
These invariants appear naturally in many areas of physics: power traces
in the theory of integrable systems on Lie algebras (Lax equation),
generalized determinants in classical field theory, and symmetrized
traces in topological field theory (Chern classes), to mention a few.

Tradition reserves different invariants for different theories, but,
since they can be viewed as symmetric functions evaluated at
eigenvalues of the given endomorphism, they are related.  In particular,
prodeterminants can be expressed as polynomials in power traces, with
the Cauchy enumeration formula giving the coefficients.

In these notes, we introduce a ``raising operator" which allows one to
construct the Cauchy coefficients in these polynomial relation.  We
study some of its properties.

Although this work is motivated by the invariants of endomorphisms, the
results are valid also for the theory of symmetric functions in general.
Sections 2 and 3 are reviews of the basic facts on prodeterminants 
and symmetric functions.  
Sections 4, 5 and 6 deal with new results.

\section{Determinants and traces of an endomorphism}  \label{s:2}

Determinant and trace are examples of invariants with respect 
to the adjoin action of the general linear group
$GL(n)$ acting on a linear space $L$, that is
$$
             \Tr gAg^{-1}=\Tr A
                   \qquad\qquad
             \det gAg^{-1}=\det A
$$
for any endomorphism $A\in\End L$ of a linear space $L$
and $g\in GL(n)$.  
These generalize into two families of invariants.  
The elements of one, called {\bf power-traces}, are defined
\begin{equation}
\label{eq:det.defI}
              I_k = {\rm Tr}\; A^k, \qquad   k=1,2,\ldots
\end{equation}
The other family generalizes determinant; its elements, denoted by $J_i$,  
are defined as coefficients of the characteristic polynomial for the
endomorphism $A$, namely
\begin{equation}
\label{eq:det.defJ}
        {\rm det}\; (A-\lambda) = \sum_{i=0}^n (-\lambda)^{n-i} \ J_i
\end{equation}
where $n=\dim L$.  We shall call $J_k$ a {\bf prodeterminant} of order $k$.
In particular
$$
\begin{array}{rcl}
               J_1(A) &=& \Tr A \quad( = I_1(A))\\
               J_n(A) &=& \det A 
\end{array}
$$
{\bf Remark:}
Generalized determinant (prodeterminant) can also be defined also as

$$
J_k = \frac{(-1)^{n-k}}{(n-k)!} \quad
      \frac{d^{n - k}}{d\lambda^{n-k}}\; 
      {\rm det}\;(A-\lambda)\big|_{\lambda = 0}
$$
or, in a more explicit form, as
\begin{equation}
\label{eq:det.defJ2}
              J_k = \sum \;  A  [a_1,a_2,\ldots ,a_k]
\end{equation}
where $A[\ldots]$ denotes determinant of the $k\times k$ minor
(submatrix) determined by the set of columns and rows indexed by the
same subset $[a_1,\ldots,a_k]$ of the set $[1,\ldots,n]$.  
The summation in (\ref{eq:det.defJ2}) extends over 
all ${n\choose k}$ possible selections of such subsets.

\bigskip\noindent
{\bf Example:}  Let $A$ be a $3\times3$ matrix
$$
  A = \left[  \begin{array}{ccc}
        a & b & c \cr
        d & e & f \cr
        g & h & i 
          \end{array}  \right] 
$$
Then one has the following prodeterminants
$$
\begin{array}{rcl}
J_1 &=&  a + e + i  \\[2pt]
J_2 &=&  \det\left[\begin{smallmatrix}a&b\cr d&e\end{smallmatrix}\right] +
         \det\left[\begin{smallmatrix}a&c\cr g&i\end{smallmatrix}\right] +
         \det\left[\begin{smallmatrix}e&f\cr h&i\end{smallmatrix}\right] \\[2pt]
J_3 &=&  \det A 
\end{array}
$$
while the power-traces are
$$
\begin{array}{rcl}
I_1  &=& a+e+i \\
I_2  &=& a^2+e^2+i^2 +2db+2gc+2fh \\
I_3  &=& a^3+e^3+i^3 +3abd+3cdg+3bde\\
     & &+ 3efg+3bfh+3fgi+3ach+3chi    
\end{array}
$$

\noindent
Particular applications in systems with symmetries
favor one of the two types of invariants,
like Lax equations use rather $I$ while the theory of Chern classes and 
some particle models --- $J$ (see Appendix A).
But, in fact, the two are dependent:
\begin{equation}
\label{eq:det.cauchy}
\begin{array}{rcl}
             J_1 &=& I_1         \cr
            2J_2 &=& I_1^2 - I_2 \cr
            6J_3 &=& I_1^3 - 3I_1I_2 + 2I_3 \cr
           12J_4 &=& I_1^4 - 6I_1^2I_2 + 8I_1I_3 +3I_2^2 - 6I_4 \cr
            \ldots&\qquad 
\end{array}
\end{equation}
The polynomials on the right side of (\ref{eq:det.cauchy}) 
will be denoted by $j$ and called {\bf Cauchy polynomials} 
--- they are the central topic of this note.
The question is to determine the coefficients of these polynomials.
They are known to be directly related to the rank of the
conjugacy classes of the symmetric group (group of permutations).
Section \ref{s:fock} will present a new formula generating the polynomials with
the corresponding coefficients.
But first, we review some basic facts about symmetric functions.

\section{Symmetric functions and Cauchy formula}  \label{s:cauchy}

Connection of the invariants with the symmetric functions follows directly 
from the fact that if an endomorphism $A$ has a diagonal matrix form 
in some basis, with eigenvalues on the diagonal
$$
A=\left[ \begin{matrix}
     \lambda_1 & 0         & \ldots &0\cr
     0         & \lambda_2 & \ldots &0\cr
     \vdots    & \vdots    & \ddots & \cr
     0         & 0         &        &\lambda_n 
           \end{matrix} \right]\, ,
$$
then clearly
$$
        I_k= \sum_i^n \; \lambda_i^k
$$
and
$$
        J_k= \sum  \; \lambda_{a_1} \cdot \lambda_{a_2}
                         \cdot  \ldots \cdot \lambda_{a_k}\, ,
$$
where the last sum runs over all $n\choose k$ selections of $k$
among the set of the $n$ eigenvalues $\lambda$ of the endomorphism $A$.
These expressions can easily be recognized as the
symmetric functions evaluated on the set of eigenvalues of $A$.
Invariants $I$ and $J$ corresponds to the two types
of basic symmetric functions.
Consequently,  relations (\ref{eq:det.cauchy}) may be understood as a transformation
formula for the change of basis of the space of symmetric functions.
\\

Let us recall some basic facts about symmetric functions.
The space of symmetric functions of degree $k$ is denoted by ${\mathcal S}_k$.
The three most frequently used families of basic symmetric
functions in variables $x_1, x_2, \ldots, x_n$ are these:

\bigskip\noindent
A.  {\caps Elementary symmetric functions:}
$$
c_k = \sum\; x_{i_1} x_{i_2} \ldots x_{i_k}
      \qquad\qquad
      1\leq i_1 < i_2 < \ldots < i_k \leq n
$$
B.  {\caps Power sums:}
\begin{equation}
\label{eq:cauchy.abc}
s_k = x_1^k + x_2^k +\ldots+ x_n^k
      \qquad\qquad
\phantom{1\leq i_1 < i_2 < \ldots < i_k \leq n}
\end{equation}
C.  {\caps Wronski functions:}
$$
w_k = \sum\; x_{i_1} x_{i_2}\ldots x_{i_k}
      \qquad\qquad
      1\leq i_1 \leq i_2 \leq \ldots \leq i_k \leq n
$$
The number $n$ of variables in these definitions may essentially be
unrestricted  --- one can always set $x_i=0$ for $i>n$.
Here is an example of the basic symmetric functions of degree 1, 2 and 3 in 
three variables:
$$
\begin{array}{rcl}
         c_1= s_1 = w_1 &=& x_1 + x_2 + x_3  \\[6pt]
         c_2 &=& x_1 x_2 + x_2 x_3 + x_1 x_3  \\
         s_2 &=& x_1^2 + x_2^2 + x_3^2        \\
         w_2 &=& x_1^2 + x_2^2 + x_3^2
              + x_1 x_2+ x_2 x_3 + x_1 x_3  \\[6pt]
         c_3 &=& x_1 x_2 x_3  \\
         s_3 &=& x_1^3 + x_2^3 + x_3^3  \\
         w_3 &=& x_1^3 + x_2^3 + x_3^3
              + x_1 x_2 x_3 + x_1^2 x_2  + x_1 x_2^2  \\
             & &+ x_1^2 x_3 + x_1 x_3^2 + x_2^2 x_3 +  x_2 x_3^2   
\end{array}
$$

\begin{definition}
By $\lambda\vdash k$ we denote a fact  that the set 
$\lambda=(\lambda_1,\lambda_2,\ldots)$ is a {\bf partition} of
natural number $k$, that is:
$$
\begin{array}{rl}
  &\lambda_1+\lambda_2+ \ldots+\lambda_k=k        \cr
  &\lambda_1\leq\lambda_2\leq\ldots\leq\lambda_k  
\end{array}
$$
A partition can alternatively be described by a {\bf partition symbol}
$\alpha=[\alpha_1,\alpha_2,\ldots,\alpha_k]$ in which $\alpha_i$ denotes
the number of {\it i}'s among the elements of set $\lambda$.  Thus
$$
     1\cdot\alpha_1+2\cdot\alpha_2+\ldots+k\cdot\alpha_k=k
$$
By $\alpha\models k$ we denote that $\alpha$ is a partition
symbol of $k$.
\end{definition}

\bigskip\noindent
{\bf Example:} Number $10$ admits partition $1+1+1+3+4=10$, thus 
we write $(1,1,1,3,4)\vdash 10$.  
For the partition symbol we write $[3,0,1,1]\models 10$.

\bigskip

The {\bf fundamental theorem of symmetric functions} states that basis of the
space of homogeneous symmetric functions of degree $k$ may be composed
from products of members of any of these families.
If $b_i$ denotes the members of any of the families of the basic functions 
(\ref{eq:cauchy.abc}), then the set of products
$$
\{\; b_1^{\alpha_1}\cdot b_2^{\alpha_2}\cdot\cdots\cdot b_k^{\alpha_k}
      \mid \alpha {\scriptstyle\models} k\;\}
$$
forms a basis of the space ${\mathcal S}_k$ of symmetric functions of degree $k$. 
The relation between the types of symmetric functions
is well known  (see, e.g., \cite{Led}).  In particular:
\\

\begin{theorem}
\label{th:fock.jjj}
Any elementary symmetric function can be expressed as
a linear combination of products of the power sums:
$$
c_k = \sum_{\alpha\models k} \;  (-1)^{\alpha_2+\alpha_4+\ldots} \;
          \cdot \frac{h(\alpha)}{n!}\;
          \cdot s_1^{\alpha_1}\cdot s_2^{\alpha_2}
          \cdot \ldots\cdot s_k^{\alpha_k}
$$
where the sum runs over all partitions of $k$ and where $h(\alpha)$
is known as {\it Cauchy formula}, is  given by
$$
     h(\alpha) = \frac{n!}{\alpha_1! \alpha_2!\cdot\cdot\cdot\alpha_k!\;\cdot\;
       1^{\alpha_1} 2^{\alpha_2}\cdots k^{\alpha_k} } \ .
$$
\end{theorem}

This theorem provides the coefficients of the polynomial expansion of
prodeterminants in terms of traces, (Eq. \ref{eq:det.cauchy}).

Interestingly, the relation between Wronski functions $w$ and 
power sums $s$ utilizes the same Cauchy formula, but without alternating 
sign:
$$
w_k = \sum_{\alpha\models k} \;  \frac{h(\alpha)}{n!}\;
          s_1^{\alpha_1}\cdot s_2^{\alpha_2}
          \cdot\ldots\cdot s_k^{\alpha_k}
$$
Proofs of this statements involve usually a rather unpleasant juggling with sums,
indices, and logarithms \cite{Led}.
In the next section, we shall give a different derivation on Cauchy
coefficients.
An alternative simple proof will be provided in Section \ref{s:graph}.
\\

\begin{remark}
\label{r:fock.comb}
Cauchy formula appears naturally in yet another context, namely in the
theory of symmetric groups.  It is well known that any
element of symmetric group $S_n$ can be written as a product of cyclic
permutations.  Two elements of $S_n$ belong to the same adjoint class if
both are composed from cycles of the same length.
From this, it immediately follows that a conjugacy class may be labeled
uniquely by a partition of $n$, say by a partition symbol
$\alpha=[\alpha_1,\alpha_2,\ldots, \alpha_n]$, and that the size of the
corresponding class is
$$
     h(\alpha) = \frac{n!}{\alpha_1!\,\alpha_2!\cdots\alpha_k!
                 \,\cdot\,
                 1^{\alpha_1} 2^{\alpha_2}\cdots k^{\alpha_k}}
$$
This is a straightforward enumeration formula.
Typically, these two appearances of Cauchy formula --- in the theory of 
symmetric functions and in symmetric group --- are left unrelated. 
Our proof of the prodeterminant-trace formula is actually 
based on the combinatorial meaning of the Cauchy formula.
\end{remark}

\section{Fock construction for Cauchy polynomials (main result)}
\label{s:fock}

In this section we provide some insight into the structure of relations
(\ref{eq:det.cauchy}) in terms a Fock space construction.  
Here we abstract from the functional meaning of the components of these 
relations, but rather we will treat
these relations as polynomials on their own.  In particular, polynomials
(\ref{eq:det.cauchy}) are obtained in a recursive process of raising a 
``vacuum'' state in the space of multivariate polynomials.
(Not to be confused with the raising operator appearing 
in a related context, like in \cite{BG}).
\\

Let ${\mathcal P}[x]$ be a linear space of finite polynomials over variables
$x_1$, $x_2$, $x_3\ldots$ etc.  We shall use the multi-index notation
$(n) = (n_1, n_2,\ldots)$. The monomials
$$
         x^{(n)} = x_1^{n_1}\; x_2^{n_2} \; x_3^{n_3} \ldots
$$
form a basis of ${\mathcal P}[x]$, and a general element of ${\mathcal P}[x]$ is
$$
    p(x) = \sum_{(n)}\;   c_{(n)}\; x^{(n)}
$$
with some coefficients $c_{(n)}$.  
We shall also use Dirac notation and write $x^{(n)}=|n_1,n_2,\ldots\rangle$, 
so that
$$
         p(x)= \sum_{(n)} c_{(n)} |n\rangle
$$
Clearly, $c_{(n)}=\langle\; n\;|\;p(x)\;\rangle$.
\\

\begin{definition}
Derivation operator $\delta \in \End{\mathcal P}[x]$ is defined 
by its action on a single variable and by Leibniz rule
\begin{equation}
\label{eq:fock.defd}
\begin{array}{rl}
  (i)\quad  &\delta \; x_k  = k\, x_{k+1}  \\
 (ii)\quad  &\delta \; (ab) = \delta a \cdot b + a\cdot \delta b 
\end{array}
\end{equation}
\end{definition}

\begin{proposition}
For a simple power $x_k^n$ and for a general monomial, 
one has respectively the following formulae
\begin{equation}
\label{eq:fock.appd}
\begin{array}{rl}
  (i)\quad  &  \delta\; x_k^n   = nk\, x_k^{n-1}\, x_{k+1}     \\[4pt]
 (ii)\quad  &  \delta\; x^{(n)} = \displaystyle\sum_{{i :\ n_i>0}} i\cdot n_i\cdot
                        x^{(n)}\cdot x_{n_{i+1}+1}\;/\;x_{n_i}  \, . 
\end{array}
\end{equation}
\end{proposition}

The last formula 
can be written in Dirac notation as
\begin{equation}
\label{eq:fock.appd2}
   \delta\; |n_1,n_2, \ldots \rangle  
           = \sum_i\;
             i\,n_i\; |n_1,n_2,\ldots,n_{i-1}, n_i-1, n_{i+1}+1, n_{i+2}, 
                \ldots\; \rangle
\end{equation}
where the sum extends over the terms for which $n_i>0$.
(Or, equivalently, one can simply set $|n_1,n_2, \ldots \rangle = 0$
whenever $n_i<0$ for some $i$).

We shall now define a {\bf raising} operator $\Delta^{-}\in\End \mathcal{P}[x]$ by
\begin{equation}
\label{eq:fock.raising}
         \Delta^{-} =  \hat x_1-\delta \ ,
\end{equation}
where $\hat x_1$ denotes operator of multiplication by variable $x_1$.
Consider a sequence of polynomials (coherent states) determined by a
consecutive application of the raising operator $\Delta^{-}$, namely
\begin{equation}
\label{eq:fock.jj}
\begin{array}{rl}
      (i)&\quad   j_1\;(x) = x_1                      \cr
     (ii)&\quad   j_{k+1}(x) = \Delta^{-} j_k = (\hat x_1-\delta)j_k \ .    
\end{array}
\end{equation}
One can easily generate the following sequence of polynomials:
\begin{equation}
\label{eq:fock.jjj}
\begin{array}{rcl}
         j_1(x) &=& x_1             \cr
         j_2(x) &=& x_1^2-x_2           \cr
         j_3(x) &=& x_1^3 -3x_1x_2 +2x_3       \cr
         j_4(x) &=& x_1^4-6x_1^2x_2+8x_1x_3+3x_2^2-6x_4      \cr
         j_5(x) &=& x_1^5-10x_1^3x_2+20x_1^2x_3+15x_1x_2^2
                                  -30x_1x_4-20x_2x_3+24x_5    \cr
%
%
         \ldots &  \quad                                
\end{array}
\end{equation}
where the first polynomial $j_1=x_1$ may be viewed as a ``vacuum state'' and
$j_k=\Delta^{k}j_1$ as the $k$-th ``excited state" obtained via the
raising operator $\Delta^{-}$.
One may easily recognize in the above the polynomials (\ref{eq:det.cauchy}):

\begin{theorem}
The system determined by raising operator $\Delta^{-}$ and Fock space
construction (\ref{eq:fock.jj}) coincides with Cauchy polynomials.  
In particular,
let $A\in\End L$ be an endomorphism of a linear space.
Denote $I_i=\Tr A^i$.  Then the $k$-th prodeterminant $J_k(A)$ is
$$
          J_k(A) = \frac{1}{k!}\; j_k(I_1, I_2, ..., I_k)
$$
\end{theorem}

A graphical proof of this theorem is provided in section \ref{s:graph}. 

\begin{remark} 
The family of polynomials $j_i$ together with
the raising operator $\Delta^{-}$ may be interpreted and studied as 
a so-called {\bf Appell system} (see \cite{FKS}).
\end{remark}

One may define a {\bf complementary raising operator}
$$
       \Delta^{+} =  \hat x_1+\delta
$$
and the corresponding sequence of polynomials
$$
\begin{array}{rl}
    (i)&\quad   k_1(x) = x_1             \cr
   (ii)&\quad   k_{i+1}(x) = (\hat x_1+\delta)k_{i} 
\end{array}
$$
Interestingly enough, the absolute values of the coefficients 
of these two types of polynomials coincide, i.e.
$$
     \langle\; j(x)\mid {\bf n}\; \rangle = \pm\; \langle\; k(x)\mid {\bf n} \;\rangle
$$
Thus $\Delta^{+}$ enumerates directly the conjugacy classes of symmetric
group:

\begin{corollary} 
The number of elements of the conjugacy class of
symmetric group $S_n$ corresponding to a partition symbol $\alpha\models
n$  (i.e., consisting of elements that are composition of cycles in
which cycle of length $i$ appears $\alpha_i$ times) is a coefficient at
$x^{(\alpha)}$ of the generating function
$$
    (x_1+\delta)^{(\alpha_1+\ldots+\alpha_k)}\;x_1
                 = \sum \; c_{(\alpha)}\; x^{(\alpha)}
$$
or, in Dirac notation,
$$
 \langle\; (x_i+\delta)^{(\alpha_1+\ldots+\alpha_k)} x_1
                         \mid \alpha_1,\ldots, \alpha_k\; \rangle
        = \sum_{|\alpha\rangle} \; c_n \; | \alpha_1,\ldots,\alpha_k \;\rangle
$$
\end{corollary}


\small

\section{Lie algebra}  \label{s:lie}

The operators  $\hat x_i$ of multiplication by $x_i$,  
the partial derivatives $\partial_i\equiv\partial/\partial x_i$
with respect to $x_i$ and identity, all acting in the space of polynomials 
${\mathcal P}[x]$, form Heisenberg Lie algebra ${\mathcal H}$, 
the algebra of the harmonic oscillator.  By including the derivation $\delta$ defined 
in (\ref{eq:fock.defd}), this algebra may be extended to a Lie algebra
\begin{equation}
\label{eq:bigH}
       \bar{\mathcal H} = \hbox{gen}\,\{\partial_i,\, \hat x_j,\, \delta\}\, .
\end{equation}
with the following commutation relations for the generators:
$$
\begin{array}{rcl}
[\; \partial_i,  \; \hat x_j\;]    &=& \delta_{ij} \cr
[\; \delta,\;       \hat x_j\;]    &=& j\cdot \hat x_{j+1} \cr
[\; \partial_j,  \; \delta\;] &=& j\cdot \hat x_{j+1} \cdot \hat x_{j}^{-2} 
\end{array}
$$
It is easy to calculate the Lie bracket of the two raising operators:
$$
    [\; \Delta^{-}, \; \Delta^{+}\;] = 2x_2\delta
$$
Consider a subspace ${\mathcal J}[x]\subset {\mathcal P}[x]$ 
spanned by polynomials $\{j_1,j_2,\ldots\}$ of Equation (\ref{eq:fock.jjj}).  
By definition, $\Delta^{-}$ is a raising operator
in the subspace ${\mathcal J}[x]$.  The derivative with
respect to the first variable acts as a lowering operator:
$$
    \partial_1 j_n = n\cdot j_{n-1}
$$
Derivative with respect to the $k$-th variable lowers the index by $k$:
$$
    \partial_k j_n = {n\choose k} \cdot j_{n-k}
$$
Recall that the number operator is an operator $N$ in the Fock space,
defined on basis elements by $N j_k=k\cdot j_k$ (eigenvectors).  In the
context of the standard Heisenberg algebra, the number operator does not 
lie in the Lie algebra, and must be defined as an element of 
the enveloping algebra, namely as a product $N = \hat x \partial$.
It is remarkable that the Lie algebra $\hat{\mathcal H}$ (\ref{eq:bigH}) does contain 
the number operator, since
$$
    [\; \partial_1,\; \Delta^{-} \;] j_n =  n \; j_n
$$
In general, one has:
\[
\begin{array}{rcl}
  [\; \Delta^{-}, \; \partial_k\;]\; j_n &=& (-)^k\;{n\choose k} \; j_{n} \\[4pt]
  [\; \Delta^{+}, \; \partial_k\;]\; k_n &=& {n\choose k} \; k_k
\end{array}
\]

\bigskip

\section{Graphic representation} \label{s:graph} 

Prodeterminants (\ref{eq:det.defJ}) can equivalently be defined 
as ``averages" over traces:
$$
J_k = \frac{1}{k!} \sum_{\sigma \in S_k}\;  \hbox{sgn\,}(\sigma)\;
                        A^{i_1}_{\sigma(i_1)} \;
                        A^{i_2}_{\sigma(i_2)} \,\cdots\,
                        A^{i_k}_{\sigma(i_k)}
$$
where an additional sum over repeated indices of terms is understood 
(Einstein's summation convention) (cf. Appendix A).  
This leads to a more geometric formulation of the algebraic objects discussed.
Here we show how a simple proof of Theorem \ref{th:fock.jjj} 
on the Fock space structure of Cauchy polynomials (Appell system)
may be obtained using a graphical language for the category of tensor spaces.
It also ties symmetric functions with the 
combinatorial meaning of the Cauchy formula of Remark \ref{r:fock.comb}. 
\\

In spirit, the graphical language for tensor contractions that we want to 
use is cognate with a number of approaches related to the language
of tensor operads like that of 
\cite{Pen},
\cite {Cvi},
\cite{Ozi1},
or \cite{Kau}.
Such a graphical language --- besides the conceptual value ---
may lead to nice simplifications of proofs, like the one we present.

\begin{center}
\begin{figure}[h]
\begin{center}
\includegraphics[width=3.5in]{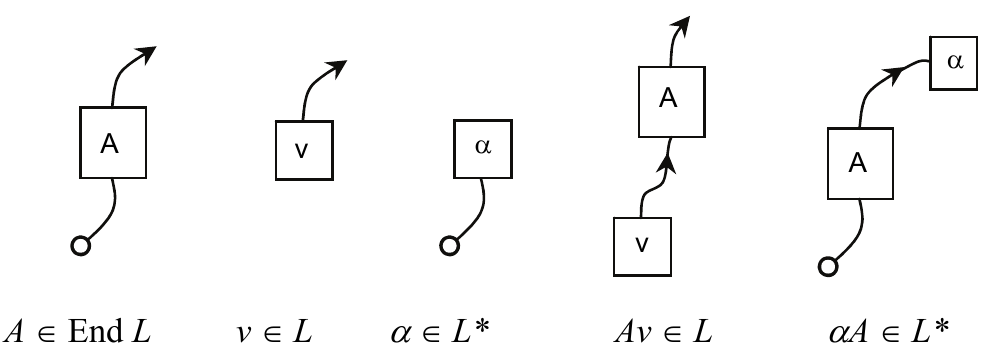}
\caption{\label{fig:contractions} Contractions of tensors}
\end{center}
\end{figure}
\end{center}

Here, we represent an endomorphism $A$ by a square with two arrows, one going out and
one going in.  The arrows may be viewed as representing indices, upper
and lower, respectively, if $A$ is represented by a matrix.  In general,
the arrows represent ``slots" of $A$ viewed as a tensor, contravariant and
covariant, respectively.  A vector will be represented by a square with
a single arrow out (``contravariant slot"); and a linear form
(covector), by a single arrow in (``covariant slot").
Figure \ref{fig:contractions} shows graphical representation 
of basic linear operations (contractions).

%
\begin{center}
\begin{figure}[h!]
\begin{center}
\includegraphics[width=3.5in]{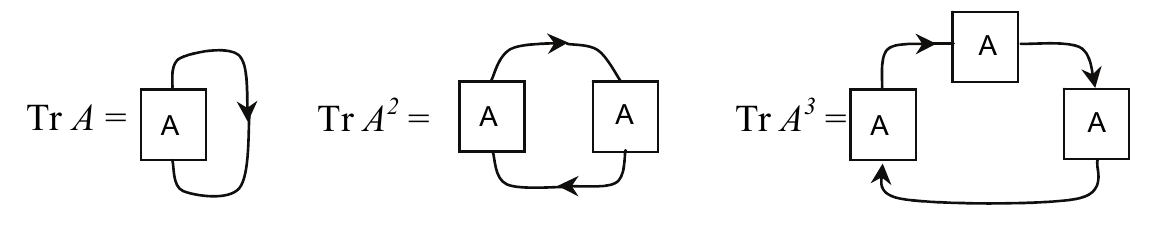}
\caption{\label{fig:traces} Power traces of an endomorphism $A$}
\end{center}
\end{figure}
\end{center}

\noindent
The trace and power-traces can be viewed as seen in Figure \ref{fig:traces}.
%

\begin{center}
\begin{figure}[h!]
\begin{center}
\includegraphics[width=3in]{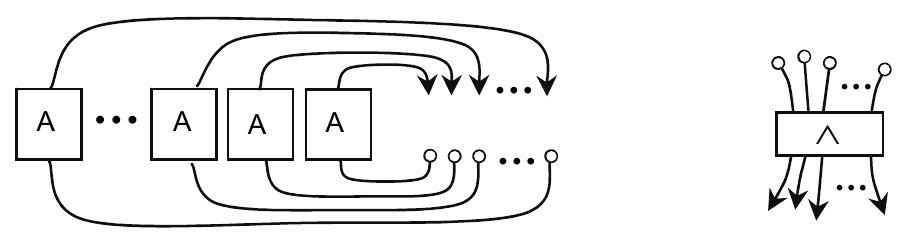}
\caption{\label{fig:tensor} Tensor product $A^{\otimes n}$ 
                             and alternating tensor $\wedge$}
\end{center}
\end{figure}
\end{center}

\noindent
The prodeterminants $J_k$ are obtained as follows.  Consider a tensor product
of $k$ copies of $A$, namely the $(k,k)$-type tensor $A\otimes A\ldots\otimes A$
(left side of Figure \ref{fig:tensor}).
One can take a trace of this operator by closing the $k$  {\it out}-arrows
with the $k$ {\it in}-arrows . 
There are $k!$ such possible pairings of the $k$ arrows with the $k$ slots.

%

\begin{figure}[H]
\begin{center}
\includegraphics[width=3.2in]{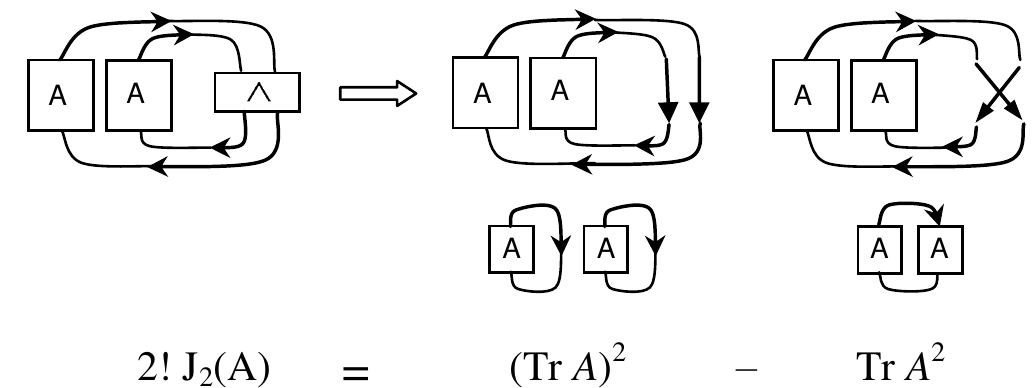}
\caption{\label{fig:case2} The origin for the second Cauchy polynomial.}
\end{center}
\end{figure}

\noindent
Scalar $J_k$ is obtained by taking a sum over all possibilities (permutations),
each term assuming the sign corresponding to the parity of the permutation.
In other words, we contract $A\otimes A\otimes\ldots\otimes A$ with 
the $(k,k)$-type alternating tensor ``$\wedge$",
totally antisymmetric in both sectors 
(represented in the right side of Figure \ref{fig:tensor}).
Figure \ref{fig:case2} illustrates the case of $J_2$.

%

\begin{figure}[h]
\begin{center}
\includegraphics[width=4.4in]{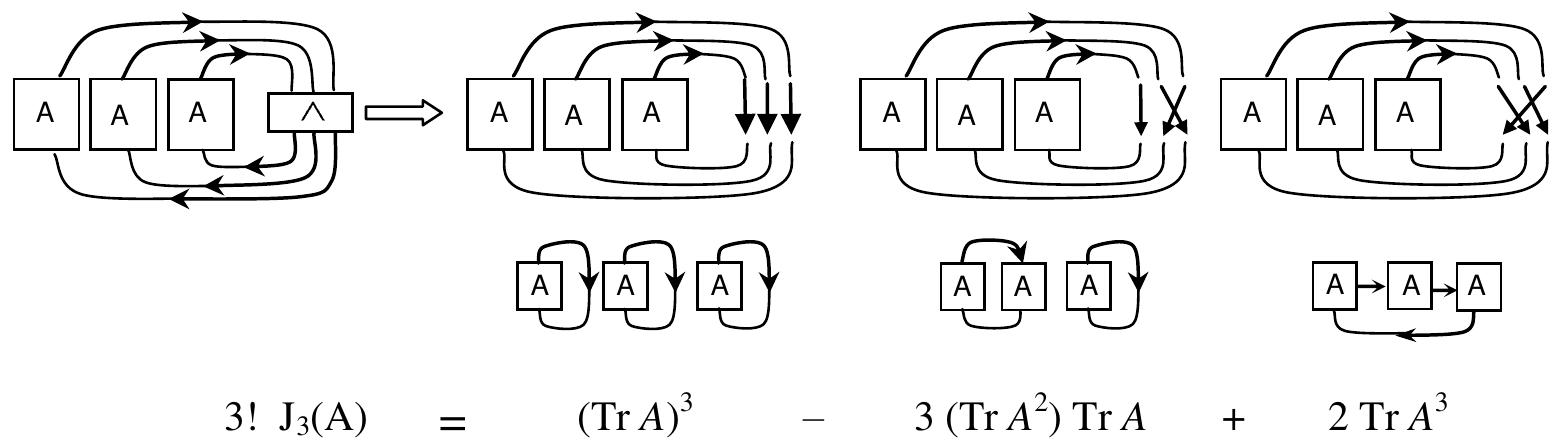}
\caption{\label{fig:case3} The origin of the third Cauchy polynomial.}
\end{center}
\end{figure}

\noindent
This gives the formula relating generalized determinants with
power-traces! In a similar simple combinatorial play with strings one obtains
the next invariant $J_3$,  see Figure \ref{fig:case3}.
%
The general form of Cauchy polynomials as a result of the 
action of the raising operator (\ref{eq:fock.raising})
emerges by induction and combinatorial meaning of the 
above tensor contractions.
Indeed, increasing the number of the ``tensor boxes" on the left side on any of the last three figures 
will add under contraction two types of new graphical loops:  either the new tensor makes a single loop with itself, 
or it will get in the path of one of the existing loops increasing their length by one.  
The former case corresponds to the operator $\hat x_1$ in (\ref{eq:fock.raising}), the latter to $\delta$.


\newpage

\section*{Appendix}  \label{s:App}

{\caps A. Geometric definition of prodeterminant} 
\\

\noindent
Given linear space $L$ of dimension $d$.  Consider Grassmann space (tensor
space)
$$
        \wedge^k L = L\wedge L\wedge\ldots\wedge L
$$
Any endomorphism $A\in \End L$ has a natural extension to an
endomorphism $A^{\wedge k}\in \End\wedge^k L$ which for a simple multi-vector
$v_1\wedge v_2\wedge\dots\wedge v_k$ is defined
$$
   A^* (v_1\wedge v_2\wedge\dots\wedge v_k) =
                        A(v_1)\wedge A(v_2)\wedge\dots\wedge A(v_k)
$$
Note that if $\{e_i\}$ is a basis of $L$, and
$\{\epsilon_i\}$ is the dual basis of $L^*$, then
$A=\epsilon^i A_i^j e_j$, and
$$
   A^{\wedge k} = A\wedge A\wedge\dots\wedge A
   = A_{i_1}^{j_1} A_{i_2}^{j_2}\ldots A_{i_k}^{j_k}
   \; e_{j_1}\wedge e_{j_2} \wedge \ldots \wedge e_{j_k}
   \otimes
   \epsilon^{i_1}\wedge \epsilon^{i_2} \wedge \ldots \wedge \epsilon^{i_k}
$$
Since multi-vectors $e_{j_1}\wedge e_{j_2} \wedge \ldots \wedge e_{j_k}$ form a
basis in the Grassmann space, the formula for prodeterminant can be
easily written as a trace of the induced endomorphism
$$
                  J_k(A)=\Tr A^{\wedge k}
$$
In particular,  $J_1(A)=\Tr A$ and $J_n(A)=\Tr A^{\wedge n}$.
\\

\noindent
{\caps B. Matrix manifolds}
\\

\noindent
Consider orbits of the adjoint action of the general linear group
$GL(n)$ acting on the space of endomorphisms $\End(L)$ over some field 
of some space $L$ of dimension $\dim L = n$:
$$
\label{eq:A1}
  g:\; M \longrightarrow gMg^{-1}
$$
The $n\times n$ dimensional space $\End\cong \mathbb{R}^{n\times n}$ is
foliated by the orbits of this action, which we call 
{\bf matrix manifolds}  (see \cite{KR}, where the case of complex spaces
is considered).
Thus the prodeterminants $J_k$ are invariant with respect to this action 
and they are natural objects to consider in this context.
In particular, a set of $n$ values determines an orbit 
(that is, the set of orbits is parameterized by the values of $J$'s).
If a matrix $M$ is an element of some orbit $\mathcal{O}_M$, then
\begin{equation}
\label{eq:A2}
       \Ker J_1(M) \cap \Ker J_2(M) \cap\ldots\cap \Ker J_n(M)
          = T_M{\mathcal O}
\end{equation}
For more (the role played by the rank of $M$ and for 
applications in field theory), see \cite{KR}.

%
%
%

\bigskip\bigskip
\noindent
{\caps Lax equation}

\bigskip\noindent
Consider a matrix representation of Lie algebra $L$ and a dynamical system
$$
\dot M = [M,B]
$$
The power-traces of $M$ provide natural invariants 
(called {\it Casimir invariants}) of the Lax dynamical system.  
Indeed
$$
\begin{array}{rcl}
  \frac{d}{dt}\;  \Tr M^n 
              &=& n \;\Tr \dot M M^{n-1} = n \Tr [M,B] M^{n-1}\cr
              &=& n (\Tr MBM^{n-1} - \Tr BM M^{n-1} ) =0  
\end{array}
\eqno(5)
$$
In particular, one can consider the adjoint representation of $L$ and
then $M$ and $B$ are directly elements of $L$. 
Let $L$ be a Lie algebra. In \cite{JK} we consider the space of $L$ as a manifold
and define a (1,1)-type tensor field (a field of endomorphisms) by
$$
         A_x (\tilde v) = (\ad_x v)^\sim= [x, v]^\sim
$$
If $x^i$ are (linear) coordinates on $L$ then
$$
         A=x^i\; c_{ij}^k \; \frac{\partial}{\partial x^k}\otimes dx^j
$$
At every point $a\in L$, tensor $A$ can be viewed as an endomorphism of
the tangent space, $A:T_aL\to T_aL$.  One can define a distribution 
$\Im A\subset TL$.  It is easy to show that this distribution is integrable; 
the integral manifolds ${\mathcal O}$ coincide with the orbits 
of the adjoint action of $L$, and:
$$
         \Im A = T{\mathcal O}
$$
The power-traces of the adjoint representation 
provide a set of scalar functions $I_i:\; L\to\mathbb{R}$.
One of them, $I_2$, is the Killing form (known in this
context as Cartan quadratic function)
$$
I_2(a)=K(a,a)=c_{ip}^q c_{qj}^p a^i a^j
$$
Constant value of the Killing function determine a pseudosphere 
(hyperbolic sphere --- in the case of semi-simple algebras).
Orbits of the adjoint action of the corresponding Lie group lie in these
spheres.  They lie inside the surfaces determined by all the
higher-order power-traces.
Thus, if for a set of $n$ numbers $r$ we define a submanifold of $L$
$$
S(r_1,\ldots,r_n) =  \{\;a\in L \mid I_1(a)=r_1,\ldots,I_n(a)=r_n\;\}
$$
then at any point $a\in L$
$$
{\mathcal O}_a \subset S(I_1(a),\ldots,I_n(a))
$$
Given vector field $B$ on $L$, define a new vector field of 
a dynamical system
$$
              X_B = A\backto B
$$
Thus we have a corollary: 
every Lax dynamical system on $L$ preserves each $I_i$ as the first integral
of motion.  Indeed:
$$
             X_B f = df \backto X_B = df \backto A\backto B  = 0
$$


\end{document}